\def\draftversion{false}
\RequirePackage{ifthen}
\ifthenelse{\equal{\draftversion}{true}}{
  \documentclass[aps,prx,10pt,galley,amsmath,amssymb,showpacs,
                 superscriptaddress]{revtex4-1}
}{
  \documentclass[aps,prx,10pt,twocolumn,showpacs,
                 superscriptaddress]{revtex4-1}
}

\usepackage{times}
\usepackage{amsmath}
\usepackage{amssymb}
\usepackage{graphicx}
\usepackage[usenames, dvipsnames]{color} 
\usepackage{bm} 
\usepackage{subfigure}

\usepackage{ulem}  

\def\Z2{$\mathbb{Z}_2$}
\def\I{\uppercase\expandafter{\romannumeral 1}}
\def\II{\uppercase\expandafter{\romannumeral 2}}
\def\III{{\uppercase\expandafter{\romannumeral 3}}}
\def\IV{{\uppercase\expandafter{\romannumeral 4}}}
\def\V{{\uppercase\expandafter{\romannumeral 5}}}

\def\angstrom{\mbox{\normalfont\AA}}

\def\kk{\mathbf{k}}

\def\00{\mathbf{k}_0,\lambda_0}

\def\crsi{CrSiTe$_3$}
\def\crge{CrGeTe$_3$}
\def\ub{$\mu_{\textrm{B}}$}

\def\Green#1{\textcolor{OliveGreen}{#1}}

\begin{document}

\title{Flux states and topological phases from spontaneous time-reversal symmetry breaking
in CrSi(Ge)Te$_3$-based systems}
\date{today}

\author{Jianpeng Liu}
\affiliation{ Kavli Institute for Theoretical Physics, University of California, Santa Barbara
CA 93106, USA}
\affiliation{ Department of Physics and Astronomy, Rutgers University,
 Piscataway, NJ 08854-8019, USA }

\author{Se Young Park}
\affiliation{ Department of Physics and Astronomy, Rutgers University,
 Piscataway, NJ 08854-8019, USA }

\author{Kevin F. Garrity}
\affiliation{Material Measurement Laboratory, National Institute of Standards and Technology, Gaithersburg MD, 20899}

\author{David Vanderbilt}
\affiliation{ Department of Physics and Astronomy, Rutgers University,
 Piscataway, NJ 08854-8019, USA }

\date{\today}

\begin{abstract}
We study adatom-covered single layers of CrSiTe$_3$ and CrGeTe$_3$
using first-principles calculations based on hybrid functionals.
We find that the insulating ground state of a monolayer of La (Lu)
deposited on single-layer \crsi\ (\crge) carries spontaneously
generated current loops around the Cr sites, These ``flux states"
induce antiferromagnetically ordered orbital moments on the
Cr sites and are also associated with nontrivial topological
properties. The calculated Chern numbers for these systems
are predicted to be $\pm 1$ even in the absence of spin-orbit
coupling, with sizable gaps on the order of 100\,meV.
The flux states and the associated topological phases result
from spontaneous time-reversal symmetry breaking due to the
presence of nonlocal Coulomb interactions.

\end{abstract}

\pacs{73.22.Gk, 75.25.Dk, 03.65.Vf}

\maketitle


\def\scr{\scriptsize}
\ifthenelse{\equal{\draftversion}{true}}{
  \marginparwidth 2.7in
  \marginparsep 0.5in
  \newcounter{comm} 
  \def\commnext{\stepcounter{comm}}
  \def\commtext{{\bf\color{blue}[\arabic{comm}]}}
  \def\commmar{{\bf\color{blue}[\arabic{comm}]}}
  \def\dvm#1{\commnext\marginpar{\small DV\commmar: #1}\commtext}
  \def\jlm#1{\commnext\marginpar{\small JPL\commmar: #1}\commtext}
  \def\sypm#1{\commnext\marginpar{\small SYP\commmar: #1}\commtext}
  \def\kgm#1{\commnext\marginpar{\small KG\commmar: #1}\commtext}
  \def\mlab#1{\marginpar{\small\bf #1}}
  \def\tnewpage{\newpage\marginpar{\small Temporary newpage}}
  \def\tfootnote#1{\Green{\scr [FOOTNOTE: #1]}}
}{
  \def\dvm#1{}
  \def\jlm#1{}
  \def\sypm#1{}
  \def\kgm#1{}
  \def\mlab#1{}
  \def\tnewpage{}
  \def\tfootnote#1{\footnote{#1}}
}



Spin-orbit coupling (SOC) has played an essential role in both
time-reversal (TR) invariant topological insulators and TR-breaking
quantum anomalous Hall (QAH) insulators
\cite{kane-rmp10,zhang-rmp11,Haldane-model,xue-science13}.  In the former, the
nontrivial
band topology typically results from band inversions
driven by SOC
\cite{kane-prl05-b,bernevig-s06,fu-prl07,zhang-np09,jpliu-spillage}.
In the latter SOC is also crucial, as it transmits the breaking of TR
symmetry from the spin sector to the orbital sector, and the breaking
of orbital TR symmetry is indispensable to obtain nonvanishing anomalous
Hall currents in insulating systems.

Recently it has been theoretically argued that topological phases may
arise even in the absence of SOC, driven only by Coulomb interactions.
For example, by studying tight-binding
models for the LaNiO$_3$/LaAlO$_3$ heterostructures
with Slater-Kanamori type local interactions, Yang \textit{et al.} and R\"uegg and Fiete
independently showed that the mean-field ground states
are in the QAH phase for certain parameters of the model even in the absence of
SOC~\cite{fiete-prb11,ran-prb11-bi}.
Moreover,
Raghu \textit{et al.}\ demonstrated that the Hartree-Fock ground state of
a tight-binding model with nonlocal
Coulomb interactions on a 2D honeycomb lattice
may be a QAH insulator, where TR symmetry is spontaneously broken due 
the nonlocal interactions \cite{raghu-prl08}.

These works suggest that
the exchange part of the Coulomb interaction is the key ingredient in
both of the aforementioned studies.  The exchange part of the multi-orbital
on-site interaction involves  the off-diagonal elements of the on-site density matrix,
which are in general complex, leading to a complex combination of
real atomic orbitals with
spontaneously generated orbital magnetic moments \cite{fiete-prb11}.
On the other hand, the exchange part of the nonlocal
interaction may give rise to a complex bond order
parameter. This acts as a complex hopping term and
generates inter-site currents \cite{raghu-prl08} like those
that arise in the Haldane model \cite{Haldane-model}.

In this work, we report a theoretical proposal for realizing
a QAH phase
driven by nonlocal Coulomb interactions in the absence of SOC in systems
based on \crsi\ and \crge\ single layers.
Unlike previous studies based on simplified lattice models
\cite{raghu-prl08,fiete-prb11,ran-prb11-bi},
we have carried out first-principles calculations
using a hybrid-functional \cite{hse03}
extension of density-functional theory (DFT) \cite{dft1, dft2}.
In the hybrid-functional approach, the exchange part of the screened
Coulomb interaction is treated as a weighted average in which a
fraction is treated using nonlocal Hartree-Fock exchange and the remainder
is calculated from a conventional semilocal-density functional.
Within the hybrid-functional approach,
we find that the ground state of a single-monolayer film of
La deposited on single-layer \crsi, or Lu on single-layer \crge, 
is a QAH insulator with an energy gap on the order of 100\,meV even in
the absence of SOC.

Our calculations show that the emergence of the topologically nontrivial
phase is accompanied by spontaneously
generated currents that flow between the Te atoms surrounding the
Cr atoms. Such a state with spontaneously generated
current loops is usually denoted as a ``flux state" or ``flux phase,''
and has been proposed as the ground state for various interacting
models~\cite{affleck-prb88,masanori-prl98, varma-prb97, varma-prl99}.
Its essential feature is that the spontaneous TR symmetry breaking
occurs in the orbital, as opposed to the spin, sector.
To the best of our knowledge, our work is the first proposal for the
appearance of a flux state and associated topological phase in a
realistic material system based on first-principles computational
methods.
Since the hybrid-functional method has been quite successful in
predicting the physical properties of a
variety of material systems, we speculate that the topologically
nontrivial flux state may in fact be the true ground state of
these systems if they can be realized in the laboratory.

\begin{figure}
\includegraphics[width=3.4in]{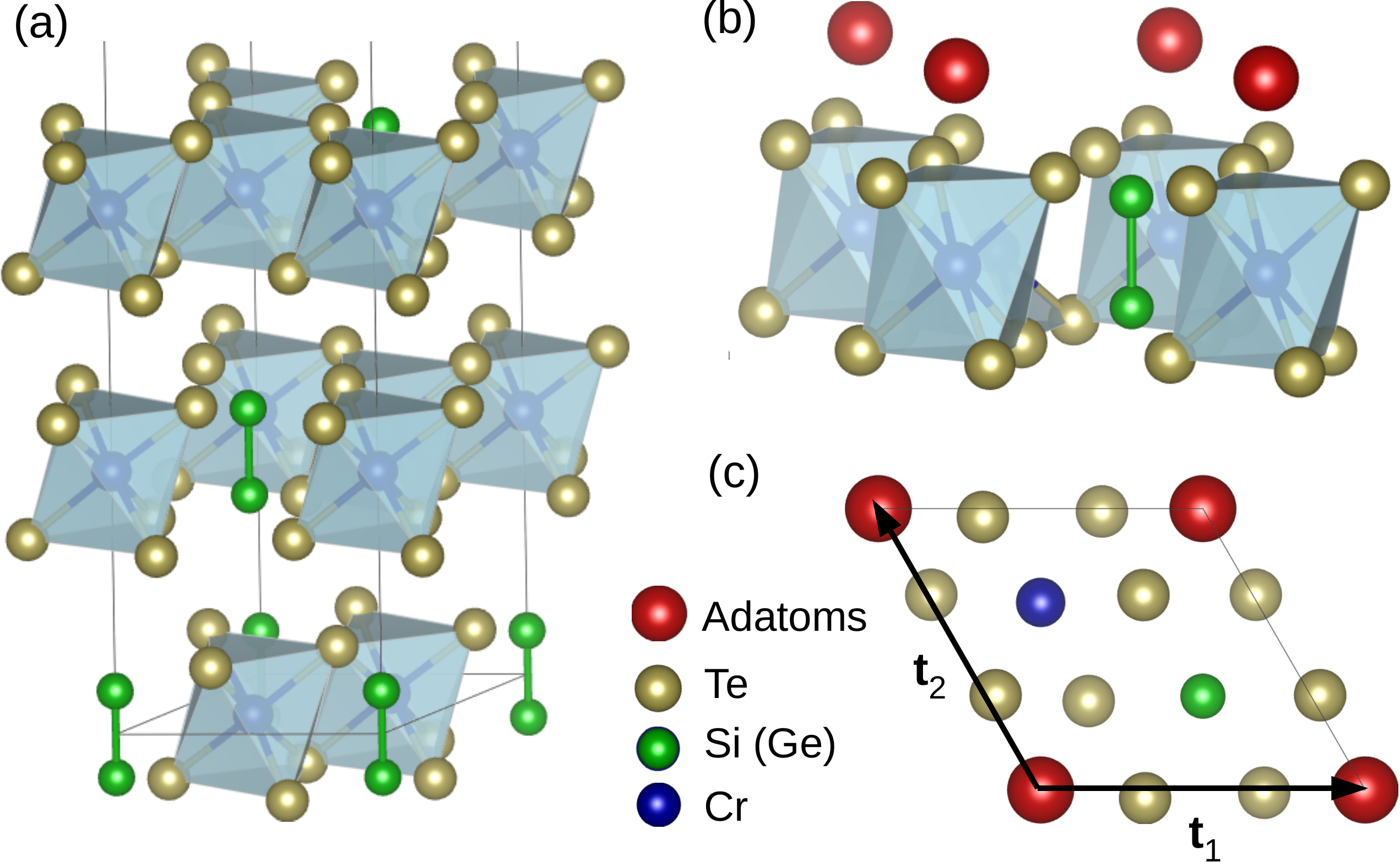}
\caption{(a) The lattice structures of bulk \crsi\ and \crge.
(b)-(c): The structure of single-layer \crsi\ (\crge) with deposited adatoms,
(b) a 3D view, and (c) a top view. $\mathbf{t}_1$ and $\mathbf{t}_2$ denote
the 2D lattice vectors of the hexagonal primitive cell.}
\label{fig:lattice}
\end{figure}

Bulk \crsi\ and \crge\ are 
ferromagnetic insulators with
Curie temperatures of 32\,K and 61\,K respectively \cite{crsite3-exp,crgete3-exp}.
As shown in Fig.~\ref{fig:lattice}(a),
the systems crystallize in a rhombohedral lattice,
forming a layered structure stacked
along the (111) 
direction with a
fairly large inter-layer spacing of $\sim$\,3.3\,\angstrom.
Each layer consists of a 2D honeycomb array of Cr atoms in
edge-sharing Te octahedra with the Si or Ge dimers inserted
into the resulting octahedral vacancy sites. The Cr
moments point normal to the layer, i.e., along the rhombohedral axis.
The weak van der Waals (vdW) inter-layer coupling
makes it easy to exfoliate thin films from bulk crystals \cite{li-crxte3}.

Recently, Garrity and Vanderbilt proposed a general strategy for
realizing the QAH state based on depositing a layer of heavy atoms
(carrying strong SOC) on the surface of an ordinary magnetic insulator
(providing TR symmetry breaking)~ \cite{garrity-prl13}.
Motivated by this proposal, we 
use first-principles calculations to study adatom layers on
\crsi\ and \crge\  (111) single layers.
As shown in Fig.~\ref{fig:lattice}(b) and (c),
we take single-layer (SL) \crsi\ or \crge\ as the substrate, and
deposit one monolayer (ML) of adatoms on top of one of the two Cr sublattices
so that they form a triangular lattice
\footnote{These two atop-CR sites are equivalent in single-layer
  \crsi\ or \crge\ by virtue of a vertical-plane mirror followed
  by TR.  We find that adatom adsorption atop either of these
  sites is energetically favored relative to the atop-dimer
  site by about 140\,meV for both materials.}

We search over a series of adatoms including
Bi, Pb, Tl, Hg, Au, Ag, In, Sb, Sn, Sc, Y, La and Lu.
The in-plane lattice constants of all the systems are fixed
at experimental values ($a$\,=\,6.773\,\angstrom\ for \crsi\ and
$a$\,=\,6.820\,\angstrom\ for \crge), but the internal atomic positions
are fully relaxed.

The structural relaxations and electronic calculations are carried out
using the Heyd-Scuseria-Ernzerhof (HSE) hybrid
functional~\cite{hse03}, which has been shown to be more
successful than traditional DFT in predicting various physical
properties such as energy gaps, lattice parameters and magnetic
moments~\cite{hse-test,hybrid-mno}.
Both the structural relaxations and electronic-structure calculations are
performed using the VASP package \cite{vasp1,vasp2}.
A $6\times 6\times 1$ $\kk$ mesh and a 380\,eV energy cutoff
are adopted, and a slab geometry is
used for all the systems.

The hybrid-functional calculations identify two
\sout{new} topological systems, namely for 1 ML La deposited on SL \crsi\ and
1 ML Lu deposited on SL \crge.  
Hereafter we will denote these two adatom systems as ``Si-La" and ``Ge-Lu" for
simplicity.

In the process of investigating the mechanism responsible for the
band inversion and topological character in these two candidate systems,
we were astonished
to find that their topological character survives
\textit{even in the absence of SOC} in the hybrid-functional framework.
This implies that the topological phase is unconventional in
that the TR symmetry is spontaneously broken directly in the spatial sector,
not by the usual SOC-mediated transmission of the TR symmetry
breaking from the spin sector to the spatial sector.
Moreover, the ground states are characterized by spontaneously
generated current flows between the bottom Te atoms surrounding the
two Cr sites in each primitive cell, leading to
antiferromagnetically ordered orbital magnetic moments on the Cr
sites.

\begin{figure}
\subfigure{
\includegraphics[width=3in]{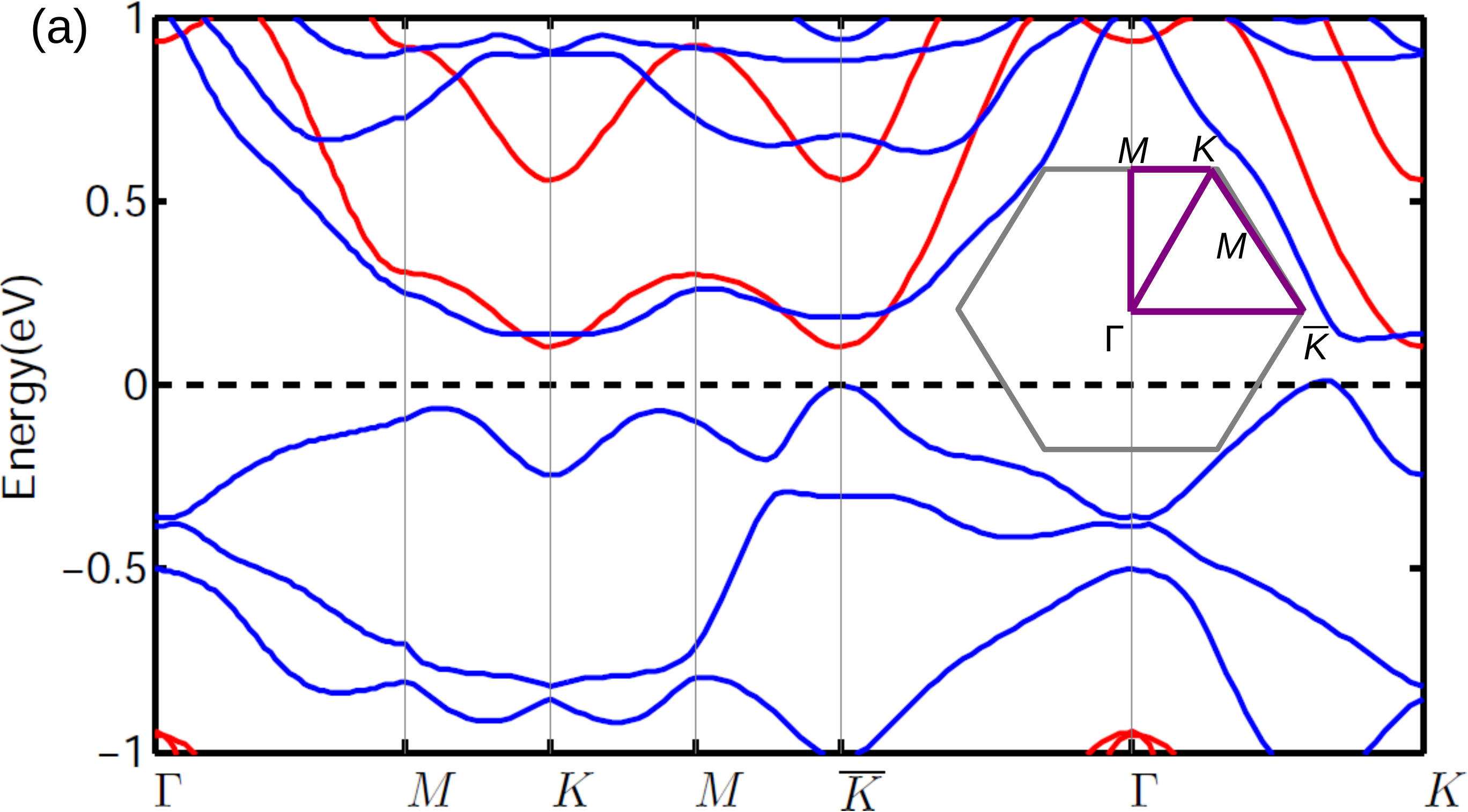}}
\subfigure{
\includegraphics[width=3in]{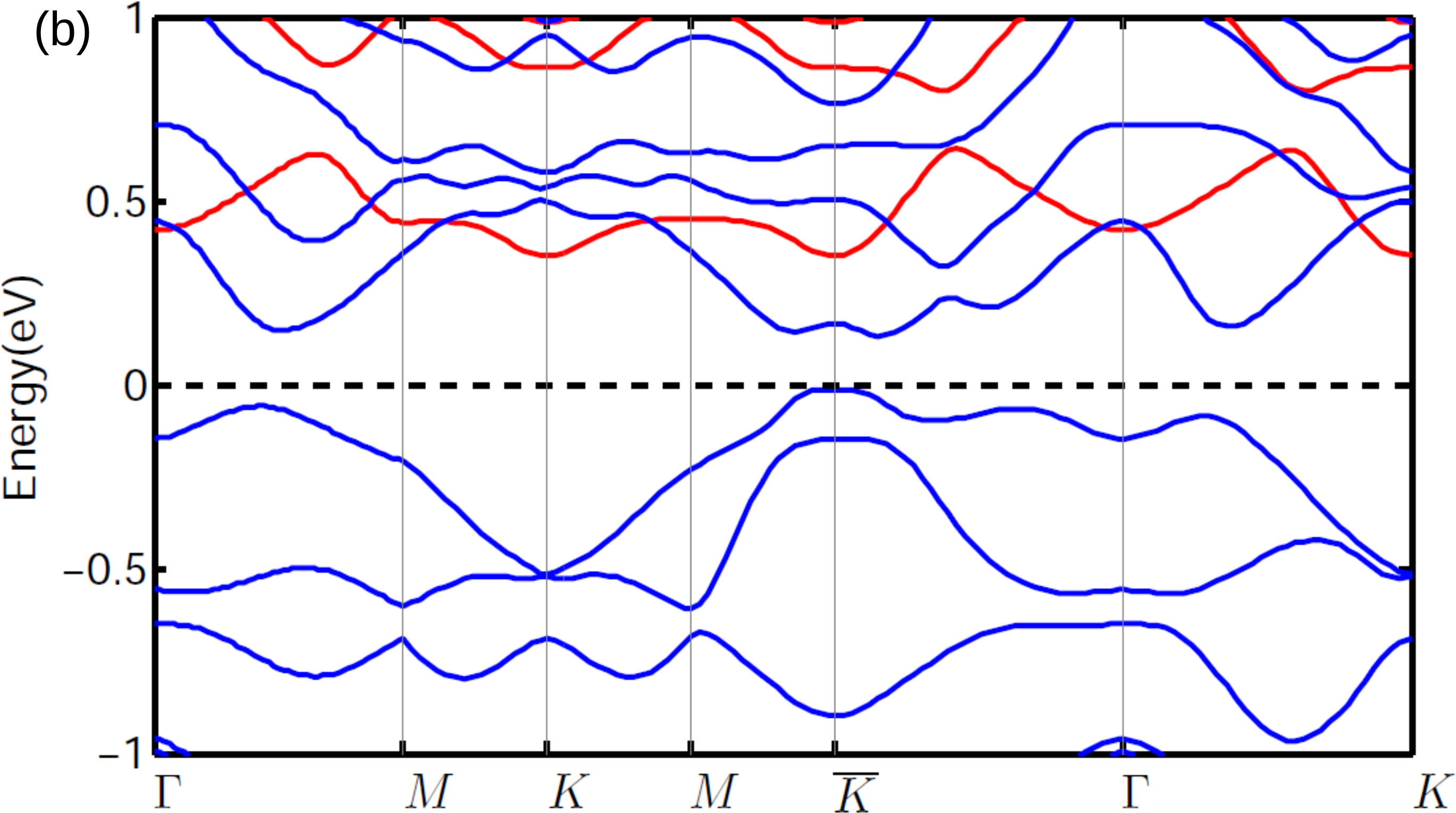}}
\caption{The bandstructures in the absence of SOC for (a) Ge-Lu,
and (b) Si-La. The red and blue curves represent
the spin majority and spin minority bands respectively. The inset in (a)
shows the Brillouin zone of a hexagonal lattice, and the bandstructures are plotted
along the high-symmetry path marked by magenta lines.}
\label{fig:band}
\end{figure}

To see how this comes about, consider
the hybrid-functional bandstructures  of Ge-Lu and Si-La
computed without SOC as shown in Figs.~\ref{fig:band} (a) and (b).
Focusing on the minority-spin (red) curves, we see that the
spatial wavefunctions obey TR symmetry, as expected when SOC is
absent; this is visible in the ``mirror symmetry'' of the curves in
panels $K$-$M$-$\bar{K}$ or $\bar{K}$-$\Gamma$-$K$, which follows
from the $\mathbf{k}\rightarrow -\mathbf{k}$ symmetry that relates
the two sides.  For the majority-spin (blue) bands, on the other
hand, this symmetry
is obviously absent. This implies that the
TR symmetry is spontaneously broken (in the orbital sector) only
in the spin-majority channel. We will discuss the interesting
properties of such ground states in the remainder of this paper.

In the \crsi\ and \crge\ single-layer systems,
the highest valence band 
and the lowest conduction band 
are mostly contributed by Te $p$ orbitals and Cr $e_g$ ($d_{yz}$ and $d_{xz}$) orbitals
respectively, all in the majority spin channel. 
If one ML of La or Lu is deposited on top of the layer,
the adatom tends to donate two
of its three valence electrons to the unoccupied Cr $e_g$ orbitals,
half-filling the four majority-spin $e_g$ bands, while the remaining
electron occupies either the $5d$ or $6s$ orbital of the adatom, also in
the majority spin channel. 
Henceforth we ignore the minority spin states.

The above analysis assumes an ionic picture in which the
hybridization between the Te $p$ and Cr $d$ orbitals is negligible.
In reality, such hybridization is quite strong in transition-metal
tellurides, as the electronegativity of Te is much weaker than that of O,
and even close to that of Cr.  Therefore,  it is more appropriate
to consider \crsi\ and \crge\ as having a strongly covalent
character.  Most of the above arguments  can be carried over
to the hybridized case, except that the orbital characters of the
four otherwise $e_g$-like states around the Fermi level become much
more complicated.  They actually consist of linear combinations of the
Cr $e_g$, Te $p$ and Cr $e_g'$ ($d_{x^2-y^2}$ and $d_{xy}$) orbitals.
In the absence of precise expressions for such complicated
hybridized orbitals, we just denote them with some simple labels
$\{\sigma_{j}, j=1,2,3,4\}$.  The extra bands from the adatom $s$
or $d$ orbitals, which also hybridize significantly with the Te $p$
orbitals, lie somewhere between the two occupied and two unoccupied
$\{{\sigma_j}\}$ bands.

Such systems with partially filled bands are expected to be metallic if
Coulomb interactions are neglected. The Fermi surface of the half-filled
$\{\sigma_j\}$ bands, however,
may be unstable against Coulomb interactions. On the other hand,
as a result of the $p$-$d$ hybridizations,
the inter-site matrix elements of the Coulomb interactions
are expected to be substantial.
Thus, on-site interactions between Cr $d$ electrons would be insufficient to
describe the effects of the Coulomb interaction in a
comprehensive manner. Therefore, it is necessary to 
adopt an \textit{ab initio} approach that takes the effects of
nonlocal Coulomb interactions into account, as is the case for
hybrid functionals.


The hybrid-functional
ground states of the two systems are found to become insulating,
leading to the  bandstructures shown in
Fig.~\ref{fig:band}.
There are a few unconventional features of
the gapped bandstructures. 
First,  as mentioned above, we note that the eigenenergies at $\mathbf{k}$
and $-\mathbf{k}$ are different 
for the majority spin,
while they are identical
for the minority spin. This indicates that
TR symmetry in the orbital sector is spontaneously broken only for the
majority spin channel.
Such TR-breaking ground states may carry
spontaneous current loops, forming flux states.
Second, there are signatures of avoided crossings around $K$ in
Fig.~\ref{fig:band}(a) and $\Gamma$ in Fig.~\ref{fig:band}(b),
suggesting that the systems may also be topologically
nontrivial. Actually if one projects the spin-majority bandstructures
for the Ge-Lu and Si-La systems onto Lu $s$ and La $d$ orbitals
respectively, one finds that there is a band-inversion character
at $\Gamma$ for Si-La and at $K$ for Ge-Lu. The projected spin-majority
bandstructures are shown in Supplementary Material.

\begin{figure}
\includegraphics[width=3in]{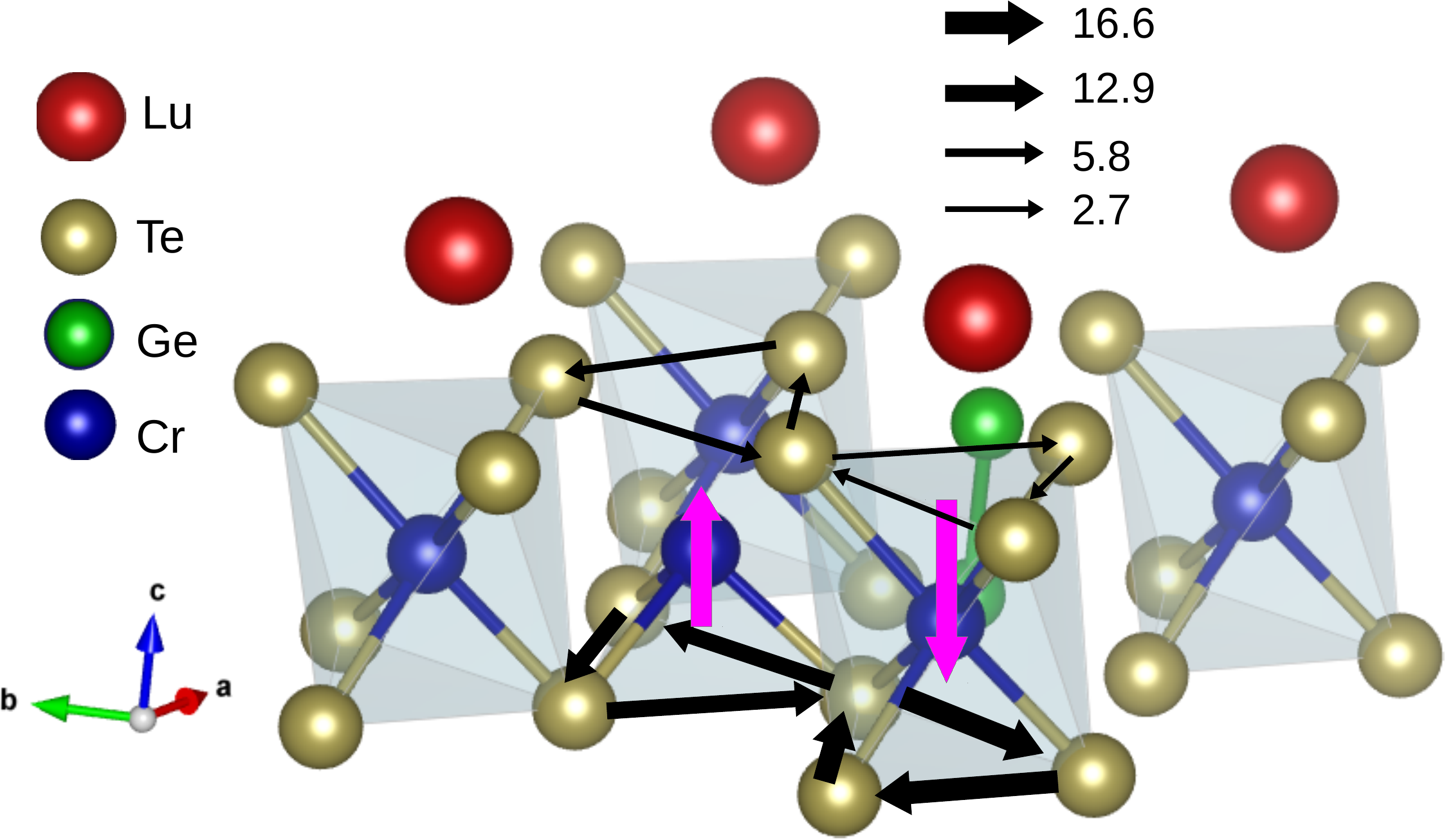}
\caption{Current loops flowing between the Te atoms
in the majority spin channel of the SOC-free Ge-Lu system.
The currents are denoted by black arrows whose thicknesses are proportional
to the magnitudes of the currents. The magenta arrows denote orbital magnetic
moments on the Cr sites. The currents are in units of 
$\mu$A.} 
\label{fig:current}
\end{figure}

To better understand the properties of these TR-breaking ground states,
we have calculated the intersite currents between atoms based on realistic
tight-binding models generated from the Wannier90
package \cite{MLWF-rmp,wannier90}
(see the Supplementary Material for details).
The calculated currents between
first-neighbor Te atoms for the Ge-Lu system,
still neglecting SOC,
are shown in Fig.~\ref{fig:current},
where the currents are represented by black arrows whose
thicknesses are proportional to the magnitudes of the currents. As is
clear from the figure, most currents flow within the bottom Te atomic
layer, forming triangular loops surrounding the Cr atoms. The two current loops
centered around the two inequivalent Cr sites circulate
in opposite directions, inducing
antiferromagnetically ordered orbital magnetic moments on the two Cr sites
as denoted by magenta arrows in Fig.~\ref{fig:current}.
Without SOC, the orbital moments
of the two Cr atoms are \{$-$0.126\,\ub , 0.113\,\ub\} for the Ge-Lu system, and
\{$-$0.066\,\ub , 0.082\,\ub\} for the Si-La system, 
with the first moment referring to that of the adatom-covered Cr site.  
Moreover, since the highest occupied band in the Si-La system is
mostly contributed by La $5d$ orbitals, there is also a relatively
large orbital moment of $-$0.18\ub\ on the La site.
For the sake of clarity, Fig.~\ref{fig:current} only depicts some of the current flows; for
example, there are considerable currents between the Si dimers and
the Te atoms, which are required to conserve the total current on each Te site.

%
Note that for each system, the configuration of orbital moments and
currents reported above is just one of two energetically equivalent
ones, since in the absence of SOC the application of spatial-only
TR (i.e., complex conjugation) will reverse all orbital moments and
currents.  We henceforth refer to the above-reported configurations
as the ``primary ones,'' and the reversed ones as ``secondary,''
even though there is nothing at this stage to prefer one over the
other.

As discussed above, there are band-inversion characters in the spin-majority
bands (see Supplementary Material),  which suggest
possible nontrivial band topologies in the two systems.
To confirm this conjecture, we calculate the Chern numbers $C$ using
the method proposed in Ref.~\onlinecite{fukui-jpsj05}, finding
$C=+1$ for the primary Ge-Lu system and $C=-1$ for the
primary Si-La system.
The calculated indirect gap based on the Wannier-interpolated
bandstructures (see the Supplementary Material) is $\sim$70\,meV for the
Ge-Lu system, and is as large as $\sim$130\,meV for the Si-La
system. Including SOC does not change the topological properties.
The indirect gap with SOC included is slightly decreased for the
Ge-Lu system ($\sim$60\,meV), while it increases to $\sim$160\,meV
for the Si-La system (see the Supplementary Material).

The primary and secondary phases, which are energetically degenerate in the
absence of SOC, are preferred by  $\sim$100\,meV over the
spatial-TR-preserving state.
With SOC included, we find that
the primary configuration is still preferred for
the Ge-Lu system, while Si-La prefers the secondary one.
Thus, both systems end up in a $C\!=\!+1$ phase.
The reversal in the Si-La system implies that the Cr and La orbital
moments all flip their signs in order to maximize the energy gain from SOC,
changing from \{$-$0.066\,\ub, 0.082\,\ub\} to
\{0.071\,\ub, $-$0.080\,\ub\} for the two Cr sites, and from
$-$0.18\,\ub\ to 0.29\,\ub\ for
the La adatom. Given that the SOC strength of La is much larger 
than that of Cr, the system evidently
selects the state with antiparallel spin and
orbital moments on the La site (the spin moment on the La site is
$-$0.644\,\ub) to maximize the energy gain from SOC.
Details of the changes when SOC is turned
on are provided in the Supplementary Material.

It should be emphasized that a flux state is not necessarily
topologically nontrivial.  That is,
there might not be any band inversion leading to a nontrivial
topology, even though TR symmetry is spontaneously broken.
We actually find this to be the case for 1 ML Lu deposited on
SL \crsi\ and 1 ML La deposited on SL \crge\ in the absence
of SOC, for which the bandstructures are presented in the
Supplementary Material.

\begin{figure}
\includegraphics[width=3.5in]{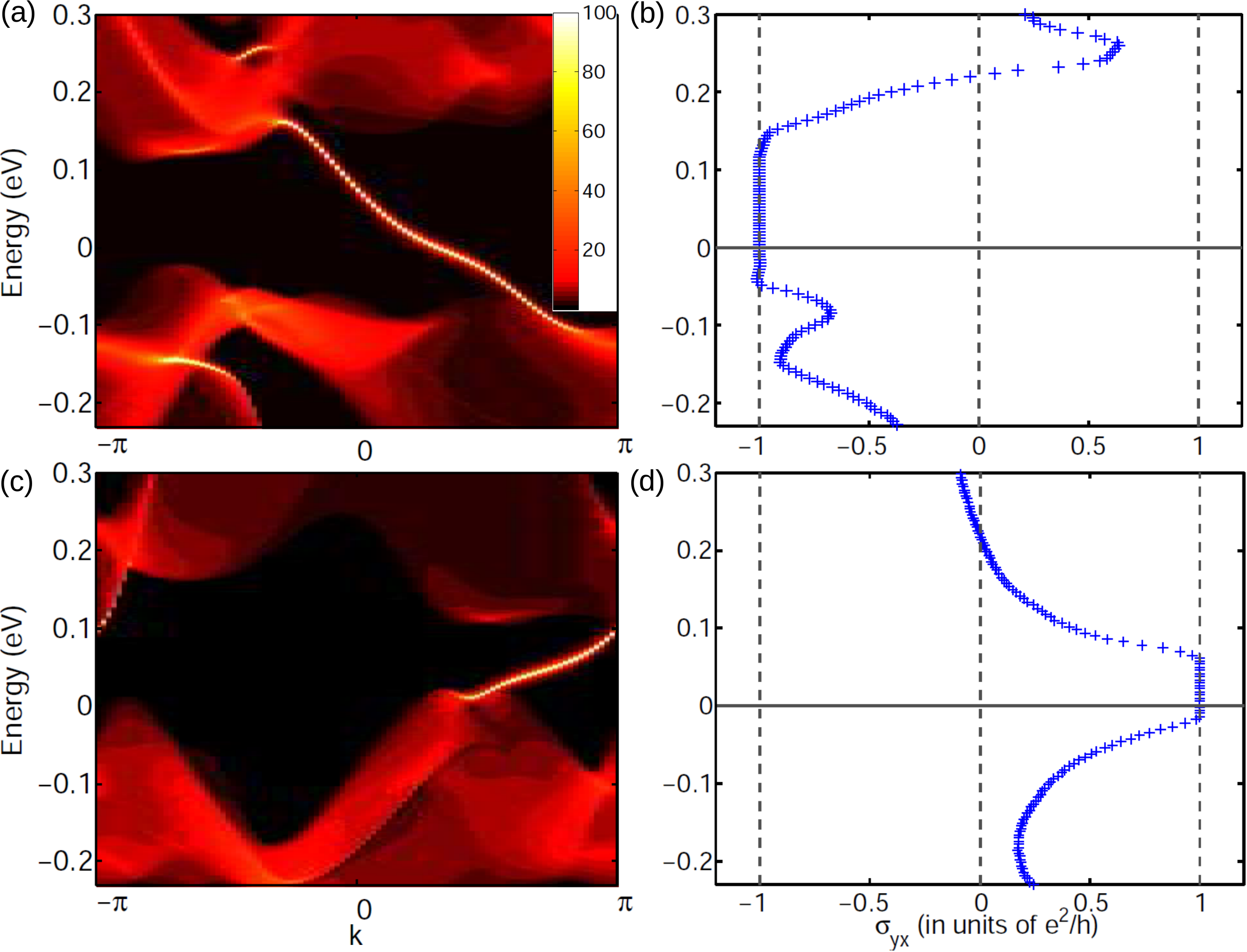}
\caption{
Left: Edge-state spectrum in the majority spin channel
for (a) Si-La and (c) Ge-Lu.
Right: Dependence of majority-spin anomalous Hall conductivities
as a function of Fermi-level position for (b) Si-La and (d) Ge-Lu.
(All without SOC.)}
\label{fig:ahc_surf}
\end{figure}

As an additional check on the computed topological character, we
have calculated the anomalous Hall conductivities $\sigma_{yx}$
and the edge states in the majority spin subspace for the primary
systems without SOC.  As shown in
Figs.~\ref{fig:ahc_surf}(a) and (c), there is a single chiral edge
state traversing through the bulk energy gap for each system. The
chiralities are opposite, since the two systems have opposite Chern
numbers in the absence of SOC.  Figs.~\ref{fig:ahc_surf}(b) and
(d) show the anomalous Hall conductances of the two systems in
the majority-spin channel as the Fermi energy is varied.
(The bulk Fermi-level positions as determined by the tetrahedron
method are set as the zero of energy in these plots.)
There are clear signatures of plateaus quantized at $\pm e^2/h$,
providing direct confirmation of the nontrivial band topology.

We now
ask whether the topological phases and the flux states
can survive if the Coulomb interactions are restricted to be local.
As mentioned above, the Coulomb interaction in the HSE hybrid
functional is nonlocal, but is screened so as to have a finite
range of the form $V(r)=(1-\textrm{erf}(r/\lambda))/r$, where
``erf" denotes the  error function and $\lambda$ is an effective
screening length~\cite{hse03}.  We have explored the behavior of
the orbital moments and Chern numbers as $\lambda$
is decreased from 5\,\angstrom\ to 1\,\angstrom.  
We characterize the staggered orbital moments by the difference
$\Delta M_{\textrm{orb}}$ between the orbital moments on the
two Cr sublattices.  While the absolute value of 
$\Delta M_{\textrm{orb}}$ starts at
0.148 (0.233)\,$\mu_\textrm{B}$ for Ge-Lu (Si-La) at $\lambda
\!=\!5$\,\AA, it falls to 0.075 (0.174)\,$\mu_\textrm{B}$ at $\lambda\!=\!1.67$\,\AA, and
0.016 (0.081)\,$\mu_\textrm{B}$ at $\lambda\!=\!1.0$\,\AA.
Both systems remain topologically nontrivial  down to
$\lambda\sim$1.3\,\angstrom, and eventually become trivial when
$\lambda\le$1\,\angstrom.  More details about the screening-length
dependence of the staggered orbital moments and the topological indices are
presented in Supplementary Material.

Interestingly, the
hybrid-functional bandstructures for $\lambda$=1\,\angstrom\ are
very similar to the DFT+$U$ bandstructures.  
We have also checked that one cannot obtain topologically nontrivial phases from
DFT+$U$ even if $U$ is tuned away from the accepted
value of 3.5\,eV.
These observations all support the conclusion that the nonlocality of
the Coulomb interactions is indispensable to obtain the observed
flux states and topological phases.

The feasibility of an experimental realization of this system
deserves some comment.  First, we have checked whether our results remain
robust for thicker \crsi\ or \crge\ layers, but unfortunately
our preliminary HSE results indicate that the flux states and
associated nontrivial topological phases do not survive even for
double-layer substrates.  This suggests that single layers of
\crsi\ or \crge\ would need to be prepared via exfoliation or
other means, presumably on an inert substrate such as silica or
boron nitride, before the adatom deposition.  Even with thicker
layers of \crsi\ or \crge, it may be possible to tune the
system using strain~\cite{crsite3-spinlattice} or chemical subsitution
in such a way as to restore the flux state.
Second, our calculations suggest that two ML of adatoms may be
energetically more stable than the single-ML configuration,
suggesting a tendency toward segregation and island
formation. Thus, low-temperature deposition of the La or Lu
monolayer may be required.  Alternatively, it may be possible to
stabilize the adatoms in monolayer form by linking them
to coordination complexes such as metallocenes \cite{crespi}.

Before concluding, we make some remarks about the
implications of our work for future 
theoretical and experimental searches for
such exchange-driven topological phases.
First, the presence of
extended and/or hybridized orbitals that can be acted upon by
nonlocal Coulomb interactions is a common feature in a wide class
of insulating materials systems.  In principle there is no need to
restrict the search to materials having ferromagnetic spin order,
or for that matter, to systems with strong SOC.  In fact, since
the scale of the gap is no longer set by the SOC strength, it may
be possible to find QAH insulators with substantially larger gaps
compared with those arising from conventional mechanisms
\cite{fang-2010science,xue-science13, garrity-prl13}.
Second, as the results reported in this paper are
obtained from a Hartree-Fock-like approximation, it is important to
inquire whether such novel ground states would survive
the application of many-body techniques beyond the mean-field level.
In view of the importance of nonlocality, we expect that it will
be necessary to go beyond approaches that focus on intrasite
correlations, such as single-site dynamical mean-field theory.
If this is not straightforward, some progress might be made
in the context of model Hamiltonians.

To summarize, we have shown that hybrid-functional calculations
predict the spontaneous breaking of orbital TR symmetry and the
emergence of flux states and associated topological phases for a
monolayer of La deposited on a single layer of \crsi, or similarly
for Lu on \crge.
We attribute the appearance of these novel phases to the exchange
component of the nonlocal Coulomb interaction acting in the
presence of strong $p$-$d$ hybridization in these transition-metal
tellurides.
The flux states are characterized by
counterpropagating current loops between the Te atoms, which
induce antiferromagnetically ordered orbital magnetic moments on
the Cr sites. The associated topological phases are characterized
by an anomalous Hall conductivity  quantized at $\pm e^2/h$ and chiral
gapless edge states even in the absence of SOC.  To the best of
our knowledge,
our work is the first proposal for a flux state
arising from spontaneous breaking of orbital TR symmetry
in a condensed-matter system
that is directly supported by first-principles calculations.
Our work is a step
forward for the understanding of topological phases 
in condensed matter physics, and may provide useful guidelines
for future experimental and theoretical works on the effects of
Coulomb interactions in transition-metal tellurides.

\acknowledgments
This work is supported by DMR-14-08838.  J. L. also would like to acknowledge
the support from DMR-15-06119.    We thank S.-W. Cheong for suggesting CrSi(Ge)Te$_3$ substrates
  as potential hosts for topological adatom structures.

\bibliography{crsite3}

\begin{thebibliography}{10}%
\makeatletter
\providecommand \@ifxundefined [1]{%
 \ifx #1\undefined \expandafter \@firstoftwo
 \else \expandafter \@secondoftwo
\fi
}%
\providecommand \@ifnum [1]{%
 \ifnum #1\expandafter \@firstoftwo
 \else \expandafter \@secondoftwo
\fi
}%
\providecommand \enquote [1]{``#1''}%
\providecommand \bibnamefont  [1]{#1}%
\providecommand \bibfnamefont [1]{#1}%
\providecommand \citenamefont [1]{#1}%
\providecommand\href[0]{\@sanitize\@href}%
\providecommand\@href[1]{\endgroup\@@startlink{#1}\endgroup\@@href}%
\providecommand\@@href[1]{#1\@@endlink}%
\providecommand \@sanitize [0]{\begingroup\catcode`\&12\catcode`\#12\relax}%
\@ifxundefined \pdfoutput {\@firstoftwo}{%
 \@ifnum{\z@=\pdfoutput}{\@firstoftwo}{\@secondoftwo}%
}{%
 \providecommand\@@startlink[1]{\leavevmode\special{html:<a href="#1">}}%
 \providecommand\@@endlink[0]{\special{html:</a>}}%
}{%
 \providecommand\@@startlink[1]{%
  \leavevmode
  \pdfstartlink
   attr{/Border[0 0 1 ]/H/I/C[0 1 1]}%
   user{/Subtype/Link/A<</Type/Action/S/URI/URI(#1)>>}%
  \relax
 }%
 \providecommand\@@endlink[0]{\pdfendlink}%
}%
\providecommand \url  [0]{\begingroup\@sanitize \@url }%
\providecommand \@url [1]{\endgroup\@href {#1}{\urlprefix}}%
\providecommand \urlprefix [0]{URL }%
\providecommand \Eprint[0]{\href }%
\@ifxundefined \urlstyle {%
  \providecommand \doi [1]{doi:\discretionary{}{}{}#1}%
}{%
  \providecommand \doi [0]{doi:\discretionary{}{}{}\begingroup
  \urlstyle{rm}\Url }%
}%
\providecommand \doibase [0]{http://dx.doi.org/}%
\providecommand \Doi[1]{\href{\doibase#1}}%
\providecommand \bibAnnote [3]{%
  \BibitemShut{#1}%
  \begin{quotation}\noindent
    \textsc{Key:}\ #2\\\textsc{Annotation:}\ #3%
  \end{quotation}%
}%
\providecommand \bibAnnoteFile [2]{%
  \IfFileExists{#2}{\bibAnnote {#1} {#2} {\input{#2}}}{}%
}%
\providecommand \typeout [0]{\immediate \write \m@ne }%
\providecommand \selectlanguage [0]{\@gobble}%
\providecommand \bibinfo [0]{\@secondoftwo}%
\providecommand \bibfield [0]{\@secondoftwo}%
\providecommand \translation [1]{[#1]}%
\providecommand \BibitemOpen[0]{}%
\providecommand \bibitemStop [0]{}%
\providecommand \bibitemNoStop [0]{.\EOS\space}%
\providecommand \EOS [0]{\spacefactor3000\relax}%
\providecommand \BibitemShut [1]{\csname bibitem#1\endcsname}%
\bibitem{kane-rmp10}%
  \BibitemOpen
  \bibfield{author}{%
  \bibinfo {author} {\bibfnamefont{M.~Z.}\ \bibnamefont{Hasan}}\ and\ \bibinfo
  {author} {\bibfnamefont{C.~L.}\ \bibnamefont{Kane}},\ }%
  \bibfield{journal}{%
  \Doi{10.1103/RevModPhys.82.3045}{\bibinfo {journal} {Rev. Mod. Phys.}}\ }%
  \textbf{\bibinfo {volume} {82}},\ \bibinfo {pages} {3045} (\bibinfo {month}
  {Nov}\ \bibinfo {year} {2010})%
  \bibAnnoteFile{NoStop}{kane-rmp10}%
\bibitem{zhang-rmp11}%
  \BibitemOpen
  \bibfield{author}{%
  \bibinfo {author} {\bibfnamefont{X.-L.}\ \bibnamefont{Qi}}\ and\ \bibinfo
  {author} {\bibfnamefont{S.-C.}\ \bibnamefont{Zhang}},\ }%
  \bibfield{journal}{%
  \Doi{10.1103/RevModPhys.83.1057}{\bibinfo {journal} {Rev. Mod. Phys.}}\ }%
  \textbf{\bibinfo {volume} {83}},\ \bibinfo {pages} {1057} (\bibinfo {month}
  {Oct}\ \bibinfo {year} {2011})%
  \bibAnnoteFile{NoStop}{zhang-rmp11}%
\bibitem{Haldane-model}%
  \BibitemOpen
  \bibfield{author}{%
  \bibinfo {author} {\bibfnamefont{F.~D.~M.}\ \bibnamefont{Haldane}},\ }%
  \bibfield{journal}{%
  \Doi{10.1103/PhysRevLett.61.2015}{\bibinfo {journal} {Phys. Rev. Lett.}}\ }%
  \textbf{\bibinfo {volume} {61}},\ \bibinfo {pages} {2015} (\bibinfo {month}
  {Oct}\ \bibinfo {year} {1988})%
  \bibAnnoteFile{NoStop}{Haldane-model}%
\bibitem{xue-science13}%
  \BibitemOpen
  \bibfield{author}{%
  \bibinfo {author} {\bibfnamefont{C.-Z.}\ \bibnamefont{Chang}}, \bibinfo
  {author} {\bibfnamefont{J.}~\bibnamefont{Zhang}}, \bibinfo {author}
  {\bibfnamefont{X.}~\bibnamefont{Feng}}, \bibinfo {author}
  {\bibfnamefont{J.}~\bibnamefont{Shen}}, \bibinfo {author}
  {\bibfnamefont{Z.}~\bibnamefont{Zhang}}, \bibinfo {author}
  {\bibfnamefont{M.}~\bibnamefont{Guo}}, \bibinfo {author}
  {\bibfnamefont{K.}~\bibnamefont{Li}}, \bibinfo {author}
  {\bibfnamefont{Y.}~\bibnamefont{Ou}}, \bibinfo {author}
  {\bibfnamefont{P.}~\bibnamefont{Wei}}, \bibinfo {author}
  {\bibfnamefont{L.-L.}\ \bibnamefont{Wang}}, \emph{et~al.},\ }%
  \bibfield{journal}{%
  \bibinfo {journal} {Science}\ }%
  \textbf{\bibinfo {volume} {340}},\ \bibinfo {pages} {167} (\bibinfo {year}
  {2013})%
  \bibAnnoteFile{NoStop}{xue-science13}%
\bibitem{kane-prl05-b}%
  \BibitemOpen
  \bibfield{author}{%
  \bibinfo {author} {\bibfnamefont{C.~L.}\ \bibnamefont{Kane}}\ and\ \bibinfo
  {author} {\bibfnamefont{E.~J.}\ \bibnamefont{Mele}},\ }%
  \bibfield{journal}{%
  \bibinfo {journal} {Phys. Rev. Lett.}\ }%
  \textbf{\bibinfo {volume} {95}},\ \bibinfo {pages} {226801} (\bibinfo {year}
  {2005})%
  \bibAnnoteFile{NoStop}{kane-prl05-b}%
\bibitem{bernevig-s06}%
  \BibitemOpen
  \bibfield{author}{%
  \bibinfo {author} {\bibfnamefont{B.~A.}\ \bibnamefont{Bernevig}}, \bibinfo
  {author} {\bibfnamefont{T.~L.}\ \bibnamefont{Hughes}},\ and\ \bibinfo
  {author} {\bibfnamefont{S.~C.}\ \bibnamefont{Zhang}},\ }%
  \bibfield{journal}{%
  \bibinfo {journal} {Science}\ }%
  \textbf{\bibinfo {volume} {314}},\ \bibinfo {pages} {1757} (\bibinfo {year}
  {2006})%
  \bibAnnoteFile{NoStop}{bernevig-s06}%
\bibitem{fu-prl07}%
  \BibitemOpen
  \bibfield{author}{%
  \bibinfo {author} {\bibfnamefont{L.}~\bibnamefont{Fu}}, \bibinfo {author}
  {\bibfnamefont{C.~L.}\ \bibnamefont{Kane}},\ and\ \bibinfo {author}
  {\bibfnamefont{E.~J.}\ \bibnamefont{Mele}},\ }%
  \bibfield{journal}{%
  \bibinfo {journal} {Phys. Rev. Lett.}\ }%
  \textbf{\bibinfo {volume} {98}},\ \bibinfo {pages} {106803} (\bibinfo {year}
  {2007})%
  \bibAnnoteFile{NoStop}{fu-prl07}%
\bibitem{zhang-np09}%
  \BibitemOpen
  \bibfield{author}{%
  \bibinfo {author} {\bibfnamefont{H.}~\bibnamefont{Zhang}}, \bibinfo {author}
  {\bibfnamefont{C.-X.}\ \bibnamefont{Liu}}, \bibinfo {author}
  {\bibfnamefont{X.-L.}\ \bibnamefont{Qi}}, \bibinfo {author}
  {\bibfnamefont{X.}~\bibnamefont{Dai}}, \bibinfo {author}
  {\bibfnamefont{Z.}~\bibnamefont{Fang}},\ and\ \bibinfo {author}
  {\bibfnamefont{S.-C.}\ \bibnamefont{Zhang}},\ }%
  \bibfield{journal}{%
  \Doi{10.1038/nphys1270}{\bibinfo {journal} {Nature Physics}}\ }%
  \textbf{\bibinfo {volume} {5}},\ \bibinfo {pages} {438} (\bibinfo {year}
  {2009})%
  \bibAnnoteFile{NoStop}{zhang-np09}%
\bibitem{jpliu-spillage}%
  \BibitemOpen
  \bibfield{author}{%
  \bibinfo {author} {\bibfnamefont{J.}~\bibnamefont{Liu}}\ and\ \bibinfo
  {author} {\bibfnamefont{D.}~\bibnamefont{Vanderbilt}},\ }%
  \bibfield{journal}{%
  \Doi{10.1103/PhysRevB.90.125133}{\bibinfo {journal} {Phys. Rev. B}}\ }%
  \textbf{\bibinfo {volume} {90}},\ \bibinfo {pages} {125133} (\bibinfo {month}
  {Sep}\ \bibinfo {year} {2014})%
  \bibAnnoteFile{NoStop}{jpliu-spillage}%
\bibitem{fiete-prb11}%
  \BibitemOpen
  \bibfield{author}{%
  \bibinfo {author} {\bibfnamefont{A.}~\bibnamefont{R\"uegg}}\ and\ \bibinfo
  {author} {\bibfnamefont{G.~A.}\ \bibnamefont{Fiete}},\ }%
  \bibfield{journal}{%
  \Doi{10.1103/PhysRevB.84.201103}{\bibinfo {journal} {Phys. Rev. B}}\ }%
  \textbf{\bibinfo {volume} {84}},\ \bibinfo {pages} {201103} (\bibinfo {month}
  {Nov}\ \bibinfo {year} {2011})%
  \bibAnnoteFile{NoStop}{fiete-prb11}%
\bibitem{ran-prb11-bi}%
  \BibitemOpen
  \bibfield{author}{%
  \bibinfo {author} {\bibfnamefont{K.-Y.}\ \bibnamefont{Yang}}, \bibinfo
  {author} {\bibfnamefont{W.}~\bibnamefont{Zhu}}, \bibinfo {author}
  {\bibfnamefont{D.}~\bibnamefont{Xiao}}, \bibinfo {author}
  {\bibfnamefont{S.}~\bibnamefont{Okamoto}}, \bibinfo {author}
  {\bibfnamefont{Z.}~\bibnamefont{Wang}},\ and\ \bibinfo {author}
  {\bibfnamefont{Y.}~\bibnamefont{Ran}},\ }%
  \bibfield{journal}{%
  \Doi{10.1103/PhysRevB.84.201104}{\bibinfo {journal} {Phys. Rev. B}}\ }%
  \textbf{\bibinfo {volume} {84}},\ \bibinfo {pages} {201104} (\bibinfo {month}
  {Nov}\ \bibinfo {year} {2011})%
  \bibAnnoteFile{NoStop}{ran-prb11-bi}%
\bibitem{raghu-prl08}%
  \BibitemOpen
  \bibfield{author}{%
  \bibinfo {author} {\bibfnamefont{S.}~\bibnamefont{Raghu}}, \bibinfo {author}
  {\bibfnamefont{X.-L.}\ \bibnamefont{Qi}}, \bibinfo {author}
  {\bibfnamefont{C.}~\bibnamefont{Honerkamp}},\ and\ \bibinfo {author}
  {\bibfnamefont{S.-C.}\ \bibnamefont{Zhang}},\ }%
  \bibfield{journal}{%
  \Doi{10.1103/PhysRevLett.100.156401}{\bibinfo {journal} {Phys. Rev. Lett.}}\
  }%
  \textbf{\bibinfo {volume} {100}},\ \bibinfo {pages} {156401} (\bibinfo
  {month} {Apr}\ \bibinfo {year} {2008})%
  \bibAnnoteFile{NoStop}{raghu-prl08}%
\bibitem{hse03}%
  \BibitemOpen
  \bibfield{author}{%
  \bibinfo {author} {\bibfnamefont{J.}~\bibnamefont{Heyd}}, \bibinfo {author}
  {\bibfnamefont{G.~E.}\ \bibnamefont{Scuseria}},\ and\ \bibinfo {author}
  {\bibfnamefont{M.}~\bibnamefont{Ernzerhof}},\ }%
  \bibfield{journal}{%
  \bibinfo {journal} {The Journal of Chemical Physics}\ }%
  \textbf{\bibinfo {volume} {118}},\ \bibinfo {pages} {8207} (\bibinfo {year}
  {2003})%
  \bibAnnoteFile{NoStop}{hse03}%
\bibitem{dft1}%
  \BibitemOpen
  \bibfield{author}{%
  \bibinfo {author} {\bibfnamefont{P.}~\bibnamefont{Hohenberg}}\ and\ \bibinfo
  {author} {\bibfnamefont{W.}~\bibnamefont{Kohn}},\ }%
  \bibfield{journal}{%
  \Doi{10.1103/PhysRev.136.B864}{\bibinfo {journal} {Phys. Rev.}}\ }%
  \textbf{\bibinfo {volume} {136}},\ \bibinfo {pages} {B864} (\bibinfo {month}
  {Nov}\ \bibinfo {year} {1964})%
  \bibAnnoteFile{NoStop}{dft1}%
\bibitem{dft2}%
  \BibitemOpen
  \bibfield{author}{%
  \bibinfo {author} {\bibfnamefont{W.}~\bibnamefont{Kohn}}\ and\ \bibinfo
  {author} {\bibfnamefont{L.~J.}\ \bibnamefont{Sham}},\ }%
  \bibfield{journal}{%
  \Doi{10.1103/PhysRev.140.A1133}{\bibinfo {journal} {Phys. Rev.}}\ }%
  \textbf{\bibinfo {volume} {140}},\ \bibinfo {pages} {A1133} (\bibinfo {month}
  {Nov}\ \bibinfo {year} {1965})%
  \bibAnnoteFile{NoStop}{dft2}%
\bibitem{affleck-prb88}%
  \BibitemOpen
  \bibfield{author}{%
  \bibinfo {author} {\bibfnamefont{I.}~\bibnamefont{Affleck}}\ and\ \bibinfo
  {author} {\bibfnamefont{J.~B.}\ \bibnamefont{Marston}},\ }%
  \bibfield{journal}{%
  \Doi{10.1103/PhysRevB.37.3774}{\bibinfo {journal} {Phys. Rev. B}}\ }%
  \textbf{\bibinfo {volume} {37}},\ \bibinfo {pages} {3774} (\bibinfo {month}
  {Mar}\ \bibinfo {year} {1988})%
  \bibAnnoteFile{NoStop}{affleck-prb88}%
\bibitem{masanori-prl98}%
  \BibitemOpen
  \bibfield{author}{%
  \bibinfo {author} {\bibfnamefont{M.}~\bibnamefont{Yamanaka}}, \bibinfo
  {author} {\bibfnamefont{W.}~\bibnamefont{Koshibae}},\ and\ \bibinfo {author}
  {\bibfnamefont{S.}~\bibnamefont{Maekawa}},\ }%
  \bibfield{journal}{%
  \Doi{10.1103/PhysRevLett.81.5604}{\bibinfo {journal} {Phys. Rev. Lett.}}\ }%
  \textbf{\bibinfo {volume} {81}},\ \bibinfo {pages} {5604} (\bibinfo {month}
  {Dec}\ \bibinfo {year} {1998})%
  \bibAnnoteFile{NoStop}{masanori-prl98}%
\bibitem{varma-prb97}%
  \BibitemOpen
  \bibfield{author}{%
  \bibinfo {author} {\bibfnamefont{C.~M.}\ \bibnamefont{Varma}},\ }%
  \bibfield{journal}{%
  \Doi{10.1103/PhysRevB.55.14554}{\bibinfo {journal} {Phys. Rev. B}}\ }%
  \textbf{\bibinfo {volume} {55}},\ \bibinfo {pages} {14554} (\bibinfo {month}
  {Jun}\ \bibinfo {year} {1997})%
  \bibAnnoteFile{NoStop}{varma-prb97}%
\bibitem{varma-prl99}%
  \BibitemOpen
  \bibfield{author}{%
  \bibinfo {author} {\bibfnamefont{C.~M.}\ \bibnamefont{Varma}},\ }%
  \bibfield{journal}{%
  \Doi{10.1103/PhysRevLett.83.3538}{\bibinfo {journal} {Phys. Rev. Lett.}}\ }%
  \textbf{\bibinfo {volume} {83}},\ \bibinfo {pages} {3538} (\bibinfo {month}
  {Oct}\ \bibinfo {year} {1999})%
  \bibAnnoteFile{NoStop}{varma-prl99}%
\bibitem{crsite3-exp}%
  \BibitemOpen
  \bibfield{author}{%
  \bibinfo {author} {\bibfnamefont{V.}~\bibnamefont{Carteaux}}, \bibinfo
  {author} {\bibfnamefont{F.}~\bibnamefont{Moussa}},\ and\ \bibinfo {author}
  {\bibfnamefont{M.}~\bibnamefont{Spiesser}},\ }%
  \bibfield{journal}{%
  \bibinfo {journal} {EPL (Europhysics Letters)}\ }%
  \textbf{\bibinfo {volume} {29}},\ \bibinfo {pages} {251} (\bibinfo {year}
  {1995})%
  \bibAnnoteFile{NoStop}{crsite3-exp}%
\bibitem{crgete3-exp}%
  \BibitemOpen
  \bibfield{author}{%
  \bibinfo {author} {\bibfnamefont{V.}~\bibnamefont{Carteaux}}, \bibinfo
  {author} {\bibfnamefont{D.}~\bibnamefont{Brunet}}, \bibinfo {author}
  {\bibfnamefont{G.}~\bibnamefont{Ouvrard}},\ and\ \bibinfo {author}
  {\bibfnamefont{G.}~\bibnamefont{Andre}},\ }%
  \bibfield{journal}{%
  \bibinfo {journal} {Journal of Physics: Condensed Matter}\ }%
  \textbf{\bibinfo {volume} {7}},\ \bibinfo {pages} {69} (\bibinfo {year}
  {1995})%
  \bibAnnoteFile{NoStop}{crgete3-exp}%
\bibitem{li-crxte3}%
  \BibitemOpen
  \bibfield{author}{%
  \bibinfo {author} {\bibfnamefont{X.}~\bibnamefont{Li}}\ and\ \bibinfo
  {author} {\bibfnamefont{J.}~\bibnamefont{Yang}},\ }%
  \bibfield{journal}{%
  \bibinfo {journal} {Journal of Materials Chemistry C}\ }%
  \textbf{\bibinfo {volume} {2}},\ \bibinfo {pages} {7071} (\bibinfo {year}
  {2014})%
  \bibAnnoteFile{NoStop}{li-crxte3}%
\bibitem{garrity-prl13}%
  \BibitemOpen
  \bibfield{author}{%
  \bibinfo {author} {\bibfnamefont{K.~F.}\ \bibnamefont{Garrity}}\ and\
  \bibinfo {author} {\bibfnamefont{D.}~\bibnamefont{Vanderbilt}},\ }%
  \bibfield{journal}{%
  \Doi{10.1103/PhysRevLett.110.116802}{\bibinfo {journal} {Phys. Rev. Lett.}}\
  }%
  \textbf{\bibinfo {volume} {110}},\ \bibinfo {pages} {116802} (\bibinfo
  {month} {Mar}\ \bibinfo {year} {2013})%
  \bibAnnoteFile{NoStop}{garrity-prl13}%
\bibitem{Note1}%
  \BibitemOpen
  \bibinfo {note} {These two atop-CR sites are equivalent in single-layer
  CrSiTe$_3$\ or CrGeTe$_3$\ by virtue of a vertical-plane mirror followed by
  TR. We find that adatom adsorption atop either of these sites is
  energetically favored relative to the atop-dimer site by about 140\protect
  \tmspace +\thinmuskip {.1667em}meV for both materials.}%
  \bibAnnoteFile{Stop}{Note1}%
\bibitem{hse-test}%
  \BibitemOpen
  \bibfield{author}{%
  \bibinfo {author} {\bibfnamefont{J.}~\bibnamefont{Heyd}}, \bibinfo {author}
  {\bibfnamefont{J.~E.}\ \bibnamefont{Peralta}}, \bibinfo {author}
  {\bibfnamefont{G.~E.}\ \bibnamefont{Scuseria}},\ and\ \bibinfo {author}
  {\bibfnamefont{R.~L.}\ \bibnamefont{Martin}},\ }%
  \bibfield{journal}{%
  \bibinfo {journal} {The Journal of chemical physics}\ }%
  \textbf{\bibinfo {volume} {123}},\ \bibinfo {pages} {174101} (\bibinfo {year}
  {2005})%
  \bibAnnoteFile{NoStop}{hse-test}%
\bibitem{hybrid-mno}%
  \BibitemOpen
  \bibfield{author}{%
  \bibinfo {author} {\bibfnamefont{C.}~\bibnamefont{Franchini}}, \bibinfo
  {author} {\bibfnamefont{R.}~\bibnamefont{Podloucky}}, \bibinfo {author}
  {\bibfnamefont{J.}~\bibnamefont{Paier}}, \bibinfo {author}
  {\bibfnamefont{M.}~\bibnamefont{Marsman}},\ and\ \bibinfo {author}
  {\bibfnamefont{G.}~\bibnamefont{Kresse}},\ }%
  \bibfield{journal}{%
  \Doi{10.1103/PhysRevB.75.195128}{\bibinfo {journal} {Phys. Rev. B}}\ }%
  \textbf{\bibinfo {volume} {75}},\ \bibinfo {pages} {195128} (\bibinfo {month}
  {May}\ \bibinfo {year} {2007})%
  \bibAnnoteFile{NoStop}{hybrid-mno}%
\bibitem{vasp1}%
  \BibitemOpen
  \bibfield{author}{%
  \bibinfo {author} {\bibfnamefont{G.}~\bibnamefont{Kresse}}\ and\ \bibinfo
  {author} {\bibfnamefont{J.}~\bibnamefont{Furthm\"uller}},\ }%
  \bibfield{journal}{%
  \Doi{10.1103/PhysRevB.54.11169}{\bibinfo {journal} {Phys. Rev. B}}\ }%
  \textbf{\bibinfo {volume} {54}},\ \bibinfo {pages} {11169} (\bibinfo {month}
  {Oct}\ \bibinfo {year} {1996})%
  \bibAnnoteFile{NoStop}{vasp1}%
\bibitem{vasp2}%
  \BibitemOpen
  \bibfield{author}{%
  \bibinfo {author} {\bibfnamefont{G.}~\bibnamefont{Kresse}}\ and\ \bibinfo
  {author} {\bibfnamefont{J.}~\bibnamefont{Furthm{\"u}ller}},\ }%
  \bibfield{journal}{%
  \bibinfo {journal} {Computational Materials Science}\ }%
  \textbf{\bibinfo {volume} {6}},\ \bibinfo {pages} {15} (\bibinfo {year}
  {1996})%
  \bibAnnoteFile{NoStop}{vasp2}%
\bibitem{MLWF-rmp}%
  \BibitemOpen
  \bibfield{author}{%
  \bibinfo {author} {\bibfnamefont{N.}~\bibnamefont{Marzari}}, \bibinfo
  {author} {\bibfnamefont{A.~A.}\ \bibnamefont{Mostofi}}, \bibinfo {author}
  {\bibfnamefont{J.~R.}\ \bibnamefont{Yates}}, \bibinfo {author}
  {\bibfnamefont{I.}~\bibnamefont{Souza}},\ and\ \bibinfo {author}
  {\bibfnamefont{D.}~\bibnamefont{Vanderbilt}},\ }%
  \bibfield{journal}{%
  \Doi{10.1103/RevModPhys.84.1419}{\bibinfo {journal} {Rev. Mod. Phys.}}\ }%
  \textbf{\bibinfo {volume} {84}},\ \bibinfo {pages} {1419} (\bibinfo {month}
  {Oct}\ \bibinfo {year} {2012})%
  \bibAnnoteFile{NoStop}{MLWF-rmp}%
\bibitem{wannier90}%
  \BibitemOpen
  \bibfield{author}{%
  \bibinfo {author} {\bibfnamefont{A.~A.}\ \bibnamefont{Mostofi}}, \bibinfo
  {author} {\bibfnamefont{J.~R.}\ \bibnamefont{Yates}}, \bibinfo {author}
  {\bibfnamefont{Y.~S.}\ \bibnamefont{Lee}}, \bibinfo {author}
  {\bibfnamefont{I.}~\bibnamefont{Souza}}, \bibinfo {author}
  {\bibfnamefont{D.}~\bibnamefont{Vanderbilt}},\ and\ \bibinfo {author}
  {\bibfnamefont{N.}~\bibnamefont{Marzari}},\ }%
  \bibfield{journal}{%
  \Doi{{10.1016/j.cpc.2007.11.016}}{\bibinfo {journal} {Computer Phys. Comm.}}\
  }%
  \textbf{\bibinfo {volume} {178}},\ \bibinfo {pages} {685} (\bibinfo {year}
  {2008})%
  \bibAnnoteFile{NoStop}{wannier90}%
\bibitem{fukui-jpsj05}%
  \BibitemOpen
  \bibfield{author}{%
  \bibinfo {author} {\bibfnamefont{T.}~\bibnamefont{Fukui}}, \bibinfo {author}
  {\bibfnamefont{Y.}~\bibnamefont{Hatsugai}},\ and\ \bibinfo {author}
  {\bibfnamefont{H.}~\bibnamefont{Suzuki}},\ }%
  \bibfield{journal}{%
  \bibinfo {journal} {Journal of the Physical Society of Japan}\ }%
  \textbf{\bibinfo {volume} {74}},\ \bibinfo {pages} {1674} (\bibinfo {year}
  {2005})%
  \bibAnnoteFile{NoStop}{fukui-jpsj05}%
\bibitem{crsite3-spinlattice}%
  \BibitemOpen
  \bibfield{author}{%
  \bibinfo {author} {\bibfnamefont{L.}~\bibnamefont{Casto}}, \bibinfo {author}
  {\bibfnamefont{A.}~\bibnamefont{Clune}}, \bibinfo {author}
  {\bibfnamefont{M.}~\bibnamefont{Yokosuk}}, \bibinfo {author}
  {\bibfnamefont{J.}~\bibnamefont{Musfeldt}}, \bibinfo {author}
  {\bibfnamefont{T.}~\bibnamefont{Williams}}, \bibinfo {author}
  {\bibfnamefont{H.}~\bibnamefont{Zhuang}}, \bibinfo {author}
  {\bibfnamefont{M.-W.}\ \bibnamefont{Lin}}, \bibinfo {author}
  {\bibfnamefont{K.}~\bibnamefont{Xiao}}, \bibinfo {author}
  {\bibfnamefont{R.}~\bibnamefont{Hennig}}, \bibinfo {author}
  {\bibfnamefont{B.}~\bibnamefont{Sales}}, \emph{et~al.},\ }%
  \bibfield{journal}{%
  \bibinfo {journal} {APL Materials}\ }%
  \textbf{\bibinfo {volume} {3}},\ \bibinfo {pages} {041515} (\bibinfo {year}
  {2015})%
  \bibAnnoteFile{NoStop}{crsite3-spinlattice}%
\bibitem{crespi}%
  \BibitemOpen
  \bibinfo {note} {V. Crespi, private communication}%
  \bibAnnoteFile{NoStop}{crespi}%
\bibitem{fang-2010science}%
  \BibitemOpen
  \bibfield{author}{%
  \bibinfo {author} {\bibfnamefont{R.}~\bibnamefont{Yu}}, \bibinfo {author}
  {\bibfnamefont{W.}~\bibnamefont{Zhang}}, \bibinfo {author}
  {\bibfnamefont{H.-J.}\ \bibnamefont{Zhang}}, \bibinfo {author}
  {\bibfnamefont{S.-C.}\ \bibnamefont{Zhang}}, \bibinfo {author}
  {\bibfnamefont{X.}~\bibnamefont{Dai}},\ and\ \bibinfo {author}
  {\bibfnamefont{Z.}~\bibnamefont{Fang}},\ }%
  \bibfield{journal}{%
  \bibinfo {journal} {Science}\ }%
  \textbf{\bibinfo {volume} {329}},\ \bibinfo {pages} {61} (\bibinfo {year}
  {2010})%
  \bibAnnoteFile{NoStop}{fang-2010science}%
\end{thebibliography}%
\end{document}